%% file: root.tex
\documentclass[letterpaper, 10 pt, conference]{ieeeconf}
\IEEEoverridecommandlockouts
\usepackage{cite}
\usepackage{amsmath,amssymb,amsfonts,amsthm}
\usepackage{algorithmic}
\usepackage{graphicx}
\usepackage{textcomp}
\usepackage{xcolor}
\usepackage{multirow}

\usepackage{caption}
\usepackage{subcaption}
\usepackage{tikz}
\usepackage{bm}
\usepackage{todonotes}

\usepackage[pscoord]{eso-pic}
\newcommand{\placetextbox}[4]{
  \setbox0=\hbox{#4}
  \AddToShipoutPictureFG*{
    \if#3r
    \put(\LenToUnit{\paperwidth-#1},\LenToUnit{\paperheight-#2}){\vtop{{\null}\makebox[0pt][r]{\begin{tabular}{r}#4\end{tabular}}}}%
    \else
    \put(\LenToUnit{#1},\LenToUnit{\paperheight-#2}){\vtop{{\null}\makebox[0pt][l]{\begin{tabular}{l}#4\end{tabular}}}}%
    \fi
  }%
}%

\newtheorem{theorem}{Theorem}

\theoremstyle{definition}
\newtheorem{definition}{Definition}

\usepackage[acronym]{glossaries}
\newacronym{EU}{EU}{European Union}
\newacronym{EV}{EV}{Electric Vehicle}
\newacronym{SOC}{SOC}{State of Charge}
\newacronym{LNW}{LNW}{Long-run Nash Welfare}
\newacronym{NW}{NW}{Nash Welfare}
\newacronym{SNE}{SNE}{Stationary Nash Equilibrium}
\newacronym{HOT}{HOT}{High Occupancy Toll}
\newacronym{HOV}{HOV}{High Occupancy Vehicle}
\newacronym[\glslongpluralkey={Markov Decision Processes}]{MDP}{MDP}{Markov Decision Process}
\newacronym{iid}{i.i.d.}{independent and identically distributed}
\newacronym{KP}{KP}{Karma Priority}
\newacronym{GP}{GP}{General Purpose}
\newacronym{DPG}{DPG}{Dynamic Population Game}
\newacronym{FCFS}{FCFS}{First-Come-First-Serve}
\newacronym{EDF}{EDF}{Earliest-Deadline-First}

\newcommand{\Real}{{\mathbb{R}}}
\newcommand{\Nat}{{\mathbb{N}}}
\newcommand{\Prob}{{\mathbb{P}}}
\newcommand{\A}{{\mathcal{A}}}
\newcommand{\C}{{\mathcal{C}}}
\newcommand{\D}{{\mathcal{D}}}
\newcommand{\N}{{\mathcal{N}}}
\newcommand{\X}{{\mathcal{X}}}
\newcommand{\T}{{\mathcal{T}}}
\newcommand{\Td}{{\mathcal{T}^\textup{d}}}

\newcommand{\U}{{\mathcal{U}}}
\newcommand{\SOC}{{\mathcal{S}}}
\newcommand{\SOCd}{\mathcal{S}^\textup d}
\newcommand{\Lot}{{\mathcal{L}}}

\newcommand{\K}{{\mathcal{K}}}
\newcommand{\B}{{\mathcal{B}}}
\newcommand{\td}{{t^\textup{d}}}
\newcommand{\sd}{{s^\textup{d}}}

\newcommand{\tstart}{{t^\textup{start}}}
\newcommand{\tend}{{t^\textup{end}}}
\newcommand{\smax}{s^\textup{max}}
\newcommand{\tdmax}{{t^\textup{d,max}}}

\newcommand{\kmax}{{k^\textup{max}}}
\newcommand{\kbar}{{\bar k}}
\newcommand{\enom}{{e^\textup{nom}}}
\newcommand{\spr}{{s^\textup{p}}}

\newcommand{\floor}[1]{\left\lfloor #1 \right\rfloor}
\newcommand{\ceil}[1]{\left\lceil #1 \right\rceil}

\DeclareMathOperator*{\argmax}{argmax}

\def\BibTeX{{\rm B\kern-.05em{\sc i\kern-.025em b}\kern-.08em
    T\kern-.1667em\lower.7ex\hbox{E}\kern-.125emX}}
\begin{document}
\bstctlcite{IEEEexample:BSTcontrol}

\title{\Large \bf
Flexible Electric Vehicle Charging with Karma
}

\author{Ezzat Elokda, Ruiting Wang, Karl H. Johansson, Angela Fontan
\thanks{The authors are with the Department of Decision and Control Systems, School of Electrical Engineering and Computer Science, KTH Royal Institute of Technology, Stockholm, Sweden (\{elokda; ruiting; kallej; angfon\}@kth.se). They are also affiliated with the research center Digital~Futures,~Stockholm,~Sweden.
This work was supported by the Wallenberg AI, Autonomous Systems and Software Program (WASP) funded by the Knut and Alice Wallenberg Foundation.}
}

\maketitle
\thispagestyle{empty}
\pagestyle{empty}

\begin{abstract}
Motivated by the need to develop fair and efficient schemes to facilitate the electrification of transport, this paper proposes a non-monetary karma economy for flexible \gls{EV} charging, managing the intertemporal allocation of limited power capacity. 
We consider a charging facility with limited capacity that must schedule arriving \glspl{EV} to charge in real-time.
For this purpose, the facility adopts online karma auctions, in which each \gls{EV} user is endowed with non-tradable karma tokens, places a karma bid in each time interval it is present in the facility, and capacity is allocated to the highest bidders, who must pay their bids.
These payments are subsequently redistributed to the users to form a closed, indefinitely sustainable economy.
The main contribution is to extend previous karma \gls{DPG} formulations to this setting which features novel \gls{SOC} dynamics and private trip deadlines in addition to urgency.
A \gls{SNE} of the \gls{EV} charging karma economy is guaranteed to exist, and it is demonstrated to provide pronounced benefits with respect to benchmark scheduling schemes as it balances between meeting deadlines and prioritizing high urgency.
\end{abstract}


\placetextbox{1.7cm}{26.3cm}{l}{\color{blue} This work has been submitted to the IEEE for possible publication. Copyright may be transferred without
notice, after \\ \color{blue} which this version may no longer be accessible.}%

\input{sections/introduction}
\input{sections/model}
\input{sections/numerical}
\input{sections/conclusion}

\bibliographystyle{IEEEtran}
\bibliography{setup,root}

\end{document}

%% file: sections/introduction.tex
\section{Introduction}

The clean energy transition is upon us.
The European Union is legally
bound to reduce carbon emissions by 55\% compared to 1990 by 2030, and achieve net-zero emissions by 2050~\cite{eu2021climatelaw}.
The electrification of transport is expected to be a major enabler of the transition owing to the presence of relatively mature, energy-efficient \acrfull{EV} technology~\cite{ec2019going}.
However, a rapid penetration of \glspl{EV} risks overloading present infrastructures if not appropriately managed.
Sources of energy scarcity facing prospective \gls{EV} users include limited charging spaces~\cite{zhang2018optimal}, physical power limits of the distribution grid~\cite{li2024impact}, and the availability of clean energy~\cite{kabir2020optimal}.
To address this challenge, it is crucial to leverage the \emph{flexibility} of \gls{EV} charging loads~\cite{sadeghianpourhamami2018quantitive}: depending on the present \acrfull{SOC} and private attributes such as upcoming trip deadline and urgency, \gls{EV} users may not always need to charge immediately when they arrive at a charging facility, making it possible to shift \gls{EV} power consumption to times of less scarcity.

The prototypical way to manage resource scarcity is \emph{monetary control}, and many monetary pricing~\cite{limmer2019dynamic,lee2020pricing,lu2022deadline} and auction~\cite{gerding2011online,hou2021simultaneous,su2025two} mechanisms have been proposed to maximize the benefit of potential flexibility in \gls{EV} charging behaviors for smart power management at charging facilities.
The working principle of these mechanisms is to expose \gls{EV} users to dynamic, real-time price signals that are high during times of scarcity, thereby incentivizing off-peak charging and other desired behaviors from the operation perspective.
However, there are several issues with these monetary control schemes.
First, they may be outright rejected in communal settings, e.g., to allocate office charging slots to company employees.
Second, empirical evidence indicates varying responsiveness to dynamic prices, with many \gls{EV} users not responding when price variations are too low to justify the burden of planning and real-time monitoring~\cite{grvzanic2022prosumers,chatzouli2025electric}.
Third and arguably most important, monetary control is \emph{discriminatory}, allocating flexibility based on willingness to pay while overlooking ability to pay~\cite{palm2026americans}.
This fairness issue is likely to exacerbate as prices vary more severely with increased \gls{EV} penetration and intermittent renewable generation.

\begin{figure}[t]
    \centering
    \includegraphics[width=\linewidth]{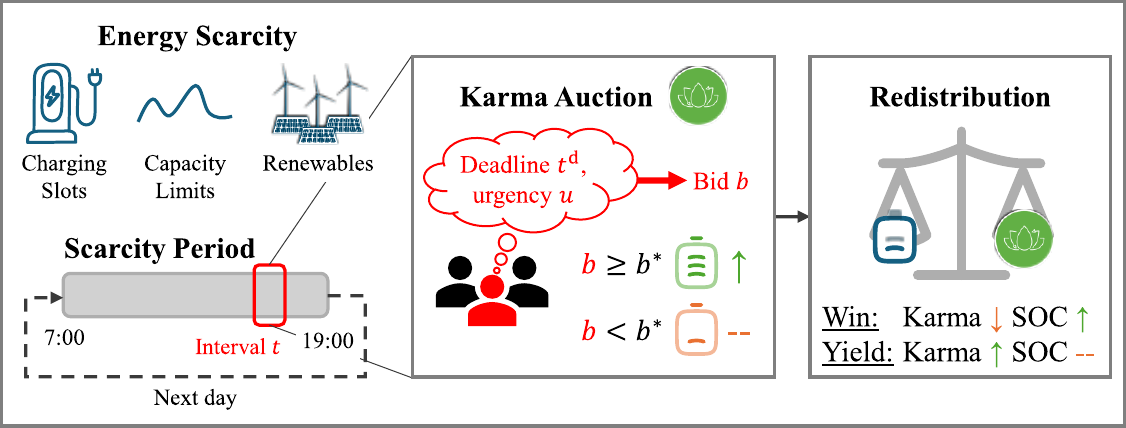}
    \caption{Illustration of \gls{EV} Charging Karma Economy.}
    \label{fig:karma-illustration}
\end{figure}

This paper instead proposes a \emph{non-monetary} approach to flexible \gls{EV} charging.
We adopt a fixed, fair charging price, and rely on a non-monetary matching mechanism to address temporal scarcity over a time period or day.
Among existing matching mechanisms~\cite{roth2015gets}, the recently developed \emph{karma economy}~\cite{elokda2024self,elokda2025carma,elokda2025vision,salazar2021urgency,pedroso2024fair} is particularly suited for this purpose due to its unique ability to be sustained indefinitely over infinite time~\cite{elokda2025vision}.
Karma is a \emph{non-tradable} token used to bid for resources repeatedly, flowing from resource consumers to yielders in a closed and indefinitely sustainable cycle.
This design encourages truthful bidding according to private needs, as users effectively ``play against their future selves''.
Compared to tradable token schemes~\cite{BALZER2023104061}, karma ensures equity through entirely non-monetary participation~\cite{elokda2025carma}.

Previous works have developed a game-theoretic framework for general karma economies~\cite{elokda2024self}, and applied it to specific problems in transportation, e.g., highway priority lanes allocation~\cite{elokda2025carma} and coupled allocation of priority roads and parking spaces~\cite{elokda2024travel}.
Moreover, it was shown that karma equilibria maximize \gls{LNW} under certain conditions~\cite{elokda2025vision}.
\gls{LNW} is the infinite-time extension of Nash welfare, an alternative social welfare function to the commonly employed utilitarianism that is considered to encode both fairness and efficiency jointly~\cite{caragiannis2019unreasonable,shilov2026welfare}.

In this paper, we conceptualize a karma-based scheme for flexible \gls{EV} charging, as portrayed in Fig.~\ref{fig:karma-illustration}.
Each day, \gls{EV} users arrive stochastically at a charging facility with private, stochastic demands to charge.
The private demands consist of a deadline to charge by, a desired charge level, and an urgency of the upcoming trip.
The charging facility employs online karma auctions, e.g., hourly, to allocate the available capacity.
In each auction, the highest karma bidders get to charge for the next hour, pay their bids, and the payments get uniformly redistributed in the population to form a closed economy.
This process repeats indefinitely, with the \gls{SOC} at the beginning of each day being correlated to the \gls{SOC} accumulated at the end of the previous day for each \gls{EV}.

In comparison to previous karma formulations~\cite{elokda2024self,elokda2025carma,elokda2024travel}, this setting bears significant complexity, owing to the presence of \emph{intra-day} charging dynamics (in addition to the previously considered \emph{inter-day} karma dynamics), and the novel scheduling aspect related to meeting private deadlines (in addition to previously considered private urgency).
Accordingly, we incorporate a micro time representing the time of day, modeled as a global state, and a macro time representing different days.
We adopt state-dependent discounting to discount future payoffs at the end of each day, a technique first used in~\cite{elokda2024travel}.
Our main contributions are as follows.
First, we adopt and extend previous karma \acrfull{DPG} formulations to this complex \gls{EV} charging problem and show that the \acrfull{SNE} existence guarantee is maintained.
Second, a numerical investigation of the \gls{SNE} reveals that the karma scheme balances effectively between meeting deadlines and prioritizing high urgency trips, according to private deadlines and urgency, and outperforms benchmark scheduling schemes, including \gls{FCFS} and \gls{EDF}, in terms of the average payoffs experienced by the users.

The remainder of the paper is organized as follows.
Section~\ref{sec:model} describes the proposed karma scheme and its \gls{DPG} formulation, in which a \gls{SNE} is guaranteed to exist (Theorem~\ref{th:SNE-exists}).
Section~\ref{sec:numerical} performs a numerical analysis of the \gls{SNE} in a charging station setting, demonstrating the benefits of the karma scheme in comparison to the benchmarks.
Section~\ref{sec:conclusion} concludes and outlines future research directions.

\paragraph*{Notation}
Let $f : \D \times \C \rightarrow \Real$ be a function whose domain is defined on $\D \subseteq \Nat$ and $\C \subseteq \Real^n$. To distinguish discrete and continuous arguments, we use notation $f[d](\bm c)$, with $d \in \D$ and $\bm{c} = (c_1,\dots,c_n) \in \C$.
We use boldface to denote vectors, matrices, and higher dimensional tensors. Accordingly, $\bm f(\bm{c}) = \left(f[d](\bm c)\right)_{d \in \D}$ is the vector formed by stacking $f[d](\bm c)$ for $d \in \D$.
We adopt the shorthand notation $\sum_d f[d]$ instead of $\sum_{d \in \D} f[d]$.
For $a \in \A \subseteq \Nat$, $\Prob[a \mid d](\bm c)$ denotes the conditional probability of $a$ given $d \in \D$ and $\bm c \in \C$, and $\Prob[d^+ \mid d](\bm c)$ denotes one-step transition probabilities for $d$.
For $m \in \Real_{\geq 0}$, we denote by $\bm \sigma \in m \: \Delta(\D):=\left\{\left. \bm \sigma \in \Real_{\geq 0}^{|\D|} \right\rvert \sum_{d \in \D} \sigma[d] = m \right\}$ a distribution over the elements of $\D$ (with total mass $m$), and $\sigma[d] \in [0, m]$ denotes the mass of element $d$.
For $c \in \Real_{\geq 0}$, $\floor{c} \in \Nat$ denotes its rounded down integer, and we define $\ceil{c} := \floor{c} + 1$.
When considering multiple user types, we denote by $x_\tau$ a quantity associated to type $\tau$.

%% file: sections/model.tex
\section{Karma-based \gls{EV} Charging}
\label{sec:model}

In this section, we first provide a high-level description of our proposed scheme in Section~\ref{sec:karma-description}, before detailing its game-theoretic model in Section~\ref{sec:dpg}.
Section~\ref{sec:SNE-exists} contains the main theoretical result on the well-posedness of the model: the guaranteed existence of a \gls{SNE} (Theorem~\ref{th:SNE-exists}).

\subsection{Description of Karma-based \gls{EV} Charging}
\label{sec:karma-description}

As illustrated in Fig.~\ref{fig:karma-illustration}, we consider a population of \gls{EV} users $\N = \{1, \dots, n\}$ who arrive at a charging facility at discrete time intervals $t \in \T = \{\tstart, \tstart + \Delta t, \dots, \tend\}$ [hr], where $\T$ is a scarcity period of interest.
The charging facility has a (potentially time-varying) capacity $C[t]$ [kWh], that is, the total energy available to charge \glspl{EV} over the interval $t$.
This could represent the available chargers in a single charging station, or more generally, the total charging capacity in a community or neighborhood of interest.
Chargers have a common nominal charging rate $\enom$ [kWh], and we assume that $C[t]$ is an integer multiple of $\enom$.

The charging facility adopts an online karma auction to allocate the available charging slots $C[t] / \enom$.
Each \gls{EV} user $i \in \N$ is endowed with \emph{karma tokens} $k_i[t] \in \K=\{0,1,\dots,\kmax\}$, and the auction proceeds over two stages: \emph{bidding} and \emph{redistribution}.
\gls{EV} users that are present at the charging facility at the start of interval $t$, and whose \gls{SOC} is less than $80\%$ of the battery capacity\footnote{We limit charging to $80\%$ of battery capacity to avoid nonlinear losses typically present in the upper range of battery charging dynamics.}, are eligible to participate in the bidding stage.
They place a \emph{karma bid} $b_i[t] \in \B[k_i[t]] = \{0,1,\dots,k_i[t]\}$, limited by their karma,
and the highest $C[t] / \enom$ bidders are admitted to charge.
Ties on the $C[t] / \enom$-highest bid are settled randomly.
In the redistribution stage, those who are admitted to charge must pay what they bid.
The total payment is redistributed uniformly to all users regardless of their presence in the charging facility~\cite{elokda2024travel}, while respecting the maximum karma constraint $\kmax$ (as elaborated in Section~\ref{par:karma-transitions}).
Thus, the system's average karma $\kbar = \sum_{i \in \N}\frac{k_i[t]}{N}$ is preserved over time. 
This process repeats indefinitely, with the karma balance at the end of each day carrying over to the next day.

Since karma constrains bids over time, bids are expected to correlate well with private user demand states, e.g., the deadline, desired \gls{SOC}, and urgency of the upcoming trip.
Users do not only compete with others but must also balance their needs against their future selves.
In what follows, we formulate a game-theoretic model that aims to predict rational equilibrium bidding behavior as a function of private demands and the resulting performance of the karma scheme.

\subsection{\acrlong{DPG} Model}
\label{sec:dpg}

Our model adopts and extends previous formulations of karma economies~\cite{elokda2024self,elokda2025carma} to the present setting, and is based on the class of \glspl{DPG}~\cite{elokda2024dynamic}.
A \gls{DPG} can be viewed as a collection of discounted, infinite-horizon \glspl{MDP} faced by individual players in a continuous population, in our case, EV users participating in online karma auctions. 
The \glspl{MDP} of different players are coupled through a \emph{social state} encoding the macroscopic distribution of states and actions in the population.
The main modeling exercise in a \gls{DPG} is to specify the \emph{immediate payoff function} and the \emph{state transition function} of these \glspl{MDP}, which are functions of the social state.
If these functions satisfy a continuity assumption, a \gls{SNE}, the central solution concept of \glspl{DPG} (defined in Section~\ref{sec:SNE-exists}), is guaranteed to exist.

Following this framework, we assume that the population $\N$ forms a continuum of nonatomic \gls{EV} users, and express the available charging capacity in average terms, denoted by $c[t]$ [kWh / user]. 
Users are assumed to be symmetric, i.e., they have identical static traits, but can be in different states at any given time\footnote{The formulation can be readily extended to multiple user types, cf.~\cite{elokda2024dynamic}.}.
The user state is defined as
\begin{subequations}
\begin{gather}
x = [t, \tilde x] \in \X = \T \times \tilde \X, \\
\tilde x = [\ell,\td,\sd,u,s,k] \in \tilde \X = \Lot \times \Td \times \SOC \times \U \times \SOC \times \K.
\end{gather}
\end{subequations}
It consists of a \emph{global} component $t \in \T$ (the current time-of-day, cf. Section~\ref{sec:karma-description}), and an \emph{individual} component $\tilde x$, which is composed as follows:
\begin{itemize}
    \item $\ell \in \Lot = \{0,1,2\}$ is a \emph{presence state}, respectively indicating that the user has not yet arrived at ($\ell=0$), is present in ($\ell=1$), or has departed from ($\ell=1$) the charging facility;
    
    \item $\td \in \Td = \{0, \Delta t, \dots, \tdmax\}$ [hr] is the user's private deadline for its upcoming trip;
    
    \item $\sd \in \SOCd \subseteq \SOC$ [kWh] is the user's private desired \gls{SOC} for its upcoming trip;
    
    \item $u \in \U = \{u^1,\dots,u^{n_u}\}$ is the user's private urgency of the upcoming trip;
    
    \item $s \in \SOC = \{\enom, 2 \, \enom, \dots, \smax\}$ [kWh] is the \gls{SOC} of the user's \gls{EV} battery, discretized by $\enom$.
    
    \item $k \in \K$ is the user's karma, cf. Section~\ref{sec:karma-description}.
\end{itemize}

The action of each user at each time-step is a karma bid $b \in \B[\ell,s,k]$, where $\B[\ell,s,k] = \emptyset$ if $\ell \neq 1$ or $s = \smax$, denoting that the user cannot participate in the bidding in these states, and $\B[\ell,s,k] = \{0,1,\dots,k\}$ otherwise.

The social state is defined as
\begin{subequations}
\begin{gather}
(\bm d, \bm \pi) \in \D \times \Pi, \\
\bm d : \T \rightarrow \Delta(\tilde \X), \quad \bm{\pi} : \X \rightarrow \Delta(\B[\ell,s,k]).
\end{gather}
\end{subequations}
The first component is the state distribution $\bm d$, defined as the conditional distribution of individual states for a given time interval, i.e., $d[\tilde x \mid t]$ is the fraction of the population in $\tilde x$ at interval $t$.
The second component is the symmetric policy $\bm \pi$ followed by all users, defined as a probabilistic map from state to action, i.e., $\pi[b \mid x]$ is the probability of bidding $b$ when in state $x$.
The remainder of this section is thus dedicated to deriving the immediate payoffs $\bm{r}(\bm d, \bm \pi)$ and the state transition probabilities $\bm{p}(\bm d, \bm \pi)$ of the users' \gls{MDP}, as functions of the social state.

\subsubsection{Immediate payoff function $r[x,b](\bm d, \bm \pi)$}
\gls{EV} users are assumed to incur a cost equal to their urgency for each time interval in which their upcoming trip deadline has passed ($\td = 0$), but they are still present in the charging facility ($\ell = 1$).
Accordingly, immediate payoffs are given by
\begin{multline}
\label{eq:immediate-rewards}
    r[x,b](\bm d, \bm \pi) = r[t,\ell,\td,\sd,s,u] = \\
    \begin{cases}
        -u, & t < \tend, \; \ell = 1, \; \td = 0, \\
        -u \left(\frac{\sd - s}{\enom} - \frac{\td}{\Delta t} \right), & t = \tend, \; \ell = 1, \frac{\td}{\Delta t} < \frac{\sd - s}{\enom}, \\
        0, & \text{otherwise}.
    \end{cases}
\end{multline}
The second case in \eqref{eq:immediate-rewards} corresponds to the situation where an \gls{EV} user has not yet completed charging at the end of the scarcity period $\tend$.
The user will continue charging at rate $\enom$ until it reaches its desired \gls{SOC} $\sd$, and incur a cost of $u$ for each additional time-step beyond its deadline.
Note that $r[x,b](\bm d, \bm \pi)$ does not actually depend on the social state $(\bm d, \bm \pi)$ nor the bid $b$, however, the resulting \gls{MDP} is still dependent on these quantities through the state transitions.

\subsubsection{State transition function $p[x^+ \mid x,b](\bm d, \bm \pi)$}
\label{sec:state-transitions}
Due to the complex definition of user states, deriving the state transition function is a significant modeling exercise that will occupy the remainder of this section.
We tackle the complexity by leveraging the conditional independence of state components to define transitions component-wise, as follows.

\paragraph{Time-of-day transitions}
Time increments by $\Delta t$ and rolls over to the next day at the end of the current day, i.e.,
\begin{align}
    \Prob[t^+ \mid t] = \begin{cases}
        1, & t^+ = t + \Delta t, \; t < \tend, \\
        1, & t^+= \tstart, \; t = \tend, \\
        0, & \textup{otherwise}.
    \end{cases}
\end{align}

\paragraph{Presence state transitions}
Given an arrival time distribution $\Prob^\textup a$ satisfying $\sum_t \Prob^\textup a[t] \leq 1$, where strict inequality models a non-zero probability to not visit the charging facility, the presence state $\ell$ evolves as
\begin{subequations}
\begin{align}
    &\Prob[\ell^+ = 1 \mid t, \ell, \td, \sd, s^+] = \notag \\
    &\qquad \begin{cases}
        \Prob^\textup a[\tstart], & t = \tend, \\[1mm]
        \frac{\Prob^\textup a[t + 
        \Delta t]}{1 - \sum_{t' \leq t} \Prob^\textup a_
        \tau[t']}, & t < \tend, \; \ell = 0, \\[1.5mm]
        1, & t < \tend, \; \ell = 1, \; \td > \Delta t, \\[1mm]
        1, & t < \tend, \; \ell = 1, \; \td \leq \Delta t, \\
        & s^+ < \sd, \\[1mm]
        0, & \textup{otherwise};
    \end{cases} \label{eq:presence-state} \\
    &\Prob[\ell^+ = 2 \mid t, \ell, \td, \sd, s^+] = \notag \\
    &\qquad \begin{cases}
        1, & t < \tend, \; \ell = 1, \; \td \leq \Delta t, \; s^+ \geq \sd, \\
        1, & t < \tend, \; \ell = 2, \\
        0, & \textup{otherwise};
    \end{cases} \label{eq:presence-state-depart}
\end{align}
\end{subequations}
and $\Prob[\ell^+ = 0 \mid \cdot] = 1 - \Prob[\ell^+ = 1 \mid \cdot] - \Prob[\ell^+ = 2 \mid \cdot]$.
The first two cases in~\eqref{eq:presence-state} capture arrivals at the charging facility, where we use Bayes rule to derive the probability of a user arriving in the next interval given that it has not arrived yet ($\ell = 0$).
Once in the charging facility $(\ell = 1)$, the user remains there until its trip deadline.
If the deadline is about to pass (or has passed already), and its next \gls{SOC} $s^+$ falls short of the desired $\sd$, it remains in the charging facility ($\ell^+ = 1$); otherwise it departs on its trip ($\ell^+ = 2$).

\paragraph{Demand state transitions}
The trip deadline, desired \gls{SOC}, and urgency jointly form the private \emph{demand state}.
They are drawn for each user at the start of each day from a distribution $\bm{\phi} \in \Delta(\Td \times \SOCd \times \U)$, and evolve according to
\begin{multline}
    \Prob[\td^+,\sd^+,u^+ \mid t, \ell,\td,\sd,u] = \\
    \begin{cases}
        \phi[\td^+,\sd^+,u^+], & t = \tend, \\[1mm]
        1, & t < \tend, \; \ell=0, \; \td^+ = \td, \\
        & \sd^+ = \sd, \; u^+ = u, \\[1mm]
        1, & t < \tend, \; \ell > 0, \\
        & \td^+ = \max\{\td - \Delta t, 0\}, \\
        & \sd^+ = \sd, \; u^+ = u, \\[1mm]
        0, & \textup{otherwise},
    \end{cases}
\end{multline}
i.e., $\sd$ and $u$ remain constant for the day, while $\td$ decrements by $\Delta t$ once the user arrives at the charging facility.

\paragraph{\gls{SOC} transitions}
The evolution of the \gls{SOC} depends on the karma auction outcomes.
Following~\cite{elokda2025carma}, let $\nu[b \mid t](\bm d, \bm \pi) = \sum_{\tilde x} d[\tilde x \mid t] \, \pi[b \mid t,\tilde x]$ denote the distribution of bids in interval $t$, from which the \emph{threshold bid} to be among the $c[t]$-highest bidders is derived as
\begin{multline}
    b^*[t](\bm d,\bm \pi) = \\
    \begin{cases}
        0, \quad \sum_{b \geq 0} \nu[b \mid t] \: \enom < c[t], \\
        \max\{b \geq 0 \mid \sum_{b' \geq b} \nu[b' \mid t] \: \enom \geq c[t]\}, \quad \textup{otherwise},
    \end{cases}
\end{multline}
where we omit the dependency on $(\bm d, \bm \pi)$ from the right-hand side of the equation (here and hereafter).
Accordingly, the energy $e \in \{0, \enom\}$ allocated to an \gls{EV} user bidding $b$ in interval $t$ is distributed as
\begin{multline}
    \label{eq:auction-outcome}
    \Prob[e=\enom \mid t,b](\bm d, \bm \pi) =
    \begin{cases}
        1, & b > b^*[t], \\
        f^\textup e[t], & b = b^*[t], \\
        0, & \textup{otherwise},
    \end{cases}
\end{multline}
where $f^\textup e[t](\bm d, \bm \pi) = \min\left\{\frac{c[t] - \sum_{b > b^*[t]} \nu[b \mid t] \, \enom}{\nu[b^*[t] \mid t] \, \enom}, 1\right\}$ is the uniform random probability for users tying at $b^*[t]$ to charge, and $\Prob[e=0 \mid \cdot] = 1 - \Prob[e=\enom \mid \cdot]$.
Note, however, that~\eqref{eq:auction-outcome} possesses a discontinuity at social states for which $\nu[b^*[t] \mid t](\bm d, \bm \pi) = 0$ for some $t$.
As in~\cite{elokda2025carma}, we approximate~\eqref{eq:auction-outcome} with a continuous function given by
\begin{multline}
    \label{eq:auction-outcome-continuous}
    \Prob^\epsilon[e=\enom \mid t,b](\bm d, \bm \pi) = \\
    \begin{cases}
        0, & \sum_{b' > b} \nu[b \mid t] \, \enom > c[t], \\
        1, & \left(\sum_{b' \geq b} \nu[b \mid t] + \epsilon\right) \enom \leq c[t], \\
        f^{\textup{e},\epsilon}[t], & \textup{otherwise},
    \end{cases}
\end{multline}
where $\epsilon$ is a small approximation parameter that slightly under-allocates $c[t]$ when there are ties at $b^*[t]$, and $f^{\textup{e},\epsilon}[t](\bm d, \bm \pi) = \min\left\{\frac{c[t] - \sum_{b > b^*[t]} \nu[b \mid t] \, \enom}{(\nu[b^*[t] \mid t] + \epsilon) \, \enom}, 1\right\}$.
It is shown in~\cite{elokda2025carma} that Equation~\eqref{eq:auction-outcome-continuous} is continuous in $(\bm d, \bm \pi)$ for $\epsilon > 0$, and $\lim_{\epsilon \rightarrow 0}\Prob^\epsilon[e \mid \cdot] = \Prob[e \mid \cdot]$.

Given the karma auction outcome probability~\eqref{eq:auction-outcome-continuous}, the \gls{SOC} transitions conditional on energy allocation $e$ are given by
\begin{subequations}
\begin{align}
    &\Prob[s^+ \mid t,\ell,\sd,s,e] = \sum_\spr \Prob[\spr \mid \ell,s,e] \, \Prob[s^+ \mid t,\sd,\spr], \\
    &\Prob[\spr \mid \ell,s,e] =
    \begin{cases}
        1, & \ell = 1, \; \spr = \min\{s + e, \smax\}, \\
        1, & \ell \neq 1, \; \spr = s, \\
        0, & \textup{otherwise},
    \end{cases} \label{eq:soc-tranistion-prior} \\
    &\Prob[s^+ \mid t,\sd,\spr] = \begin{cases}
        1, & t < \tend, \; s^+ = \spr, \\
        \Prob^\textup{end}[s^+ \mid \sd, \spr], & t = \tend, \\
        0, & \textup{otherwise},
    \end{cases} \label{eq:soc-tranistion-end}
\end{align}
\end{subequations}
where $\spr$ is an intermediate variable denoting the \gls{SOC} prior to completing the upcoming trip, distributed per~\eqref{eq:soc-tranistion-prior}.
The \gls{SOC} transitions conditional on $\spr$ are given by~\eqref{eq:soc-tranistion-end}, in which $\Prob^\textup{end}[s^\textup{start} \mid \sd, s^\textup{end}]$ models an exogenous distribution relating the current day's $\sd$ and ending \gls{SOC} $s^\textup{end}$ to the next day's starting \gls{SOC} $s^\textup{start}$, cf. Section~\ref{sec:params} for an example.

\paragraph{Karma transitions}
\label{par:karma-transitions}
It remains to specify the karma transitions which follow the same redistributive scheme of~\cite{elokda2025carma}.
Consider as intermediate variable the next karma prior to redistribution $k^\textup p$, governed by
\begin{align}
\label{eq:prekarma-transitions}
    \Prob[k^\textup p \mid k, b, e] =
    \begin{cases}
        1, & e = 0, \; k^\textup p = k \\
        1, & e = \enom, \; k^\textup p = k - b \\
        0, & \textup{otherwise}.
    \end{cases}
\end{align}
The distribution of $k^\textup p$ in interval $t$ is
$\Prob[k^\textup p \mid t](\bm d,\bm \pi) = \sum_{\tau,\tilde x} d[\tilde x] \sum_b \pi[b \mid t,\tilde x] \sum_e \Prob^\epsilon[e \mid t,b] \, \Prob[k^\textup p \mid k,b,e]$.
The uniform redistribution share to each user is thus computed as $\bar \xi[t](\bm d,\bm \pi) = \bar k - \sum_{k^\textup p} \Prob[k^\textup p \mid t] \, k^\textup p$.
The following karma-dependent redistribution attains $\bar \xi[t]$ on average, meanwhile respecting the maximum karma limit $\kmax$:
\begin{multline}
    \xi[t,k^\textup p](\bm d,\bm \pi) = \\
    \begin{cases}
    \kmax - k^\textup p, & k^\textup p + \floor{\bar \xi[t]} \geq \kmax \\
    \frac{\bar \xi[t] - \sum_{k^{\textup p'} \geq \kmax - \floor{\bar \xi[t]}} \Prob[k^{\textup p'} \mid t] \left(\kmax - k^{\textup p'}\right)}{\sum_{k^{\textup p'} < \kmax - \floor{\bar \xi[t]}} \Prob[k^{\textup p'} \mid t]}, & \textup{otherwise}. \end{cases}
\end{multline}
Letting $f^k[t,k^\textup p](\bm d,\bm \pi) = \xi[t,k^\textup p] - \floor{\xi[t,k^\textup p]}$, the karma transitions conditional on $k^\textup{p}$ are
\begin{multline}
\label{eq:karma-transitions-prekarma}
    \Prob[k^+ \mid t,k^\textup p](\bm d, \bm \pi) = \\
    \begin{cases} 1 - f^k[t,k^\textup p], & k^+ = k^\textup p + \floor{\xi[t,k^\textup p]}, \\
    f^k[t,k^\textup p], & k^+ = k^\textup p + \ceil{\xi[t,k^\textup p]}, \\
    0, & \textup{otherwise},
    \end{cases}
\end{multline}
and the overall karma transitions are obtained by combining~\eqref{eq:prekarma-transitions}--\eqref{eq:karma-transitions-prekarma} as
\begin{multline}
    \Prob[k^+ \mid t,k,b,e](\bm d,\bm \pi) = \sum_{k^\textup p} \Prob[k^\textup p \mid k,b,e] \, \Prob[k^+ \mid t,k^\textup p].
\end{multline}

\paragraph{Overall state transitions}
Finally, combining the above gives the overall state transition function
\begin{multline}
    \label{eq:state-transitions}
    p[x^+ \mid x, b](\bm d,\bm \pi) = \Prob[t^+ \mid t] \\
    \Prob[\td^+,\sd^+,u^+ \mid t,\ell,\td,\sd,u] \\
    \Prob[\ell^+,s^+,k^+\mid t,\ell,\td,\sd,s,k,b],
\end{multline}
where the $[\ell,s,k]$-state transitions are grouped using
\begin{multline}
    \Prob[\ell^+,s^+,k^+\mid t,\ell,\td,\sd,s,k,b] = \\
    \sum_e \Prob^\epsilon[e \mid t,b] \, \Prob[s^+ \mid t,\ell,\sd,s,e] \\
    \Prob[\ell^+ = 1 \mid t, \ell, \td, \sd, s^+] \, \Prob[k^+ \mid t,k,b,e].
\end{multline}

\subsection{Existence of \acrfull{SNE}}
\label{sec:SNE-exists}
With the immediate payoffs~\eqref{eq:immediate-rewards} and state transitions~\eqref{eq:state-transitions} specified, we proceed to define the solution concept of \gls{SNE} and guarantee its existence in accordance to the \gls{DPG} framework.
The state transition matrix $\bm{P}(\bm d,\bm \pi)$, infinite horizon value function $\bm{V}(\bm d,\bm \pi)$, and state-action value function $\bm{Q}(\bm d,\bm \pi)$ are respectively defined as
\begin{subequations}
\begin{align*}
    &P[x^+ \mid x] = \sum_b \pi[b \mid x] \: p[x^+ \mid x, b], \\
    &V[x] = r[t,\ell,\td,\sd,s,u] + \delta[t] \sum_{x^+} P[x^+ \mid x] \: V[x^+], \\
    &Q[x,b] = r[t,\ell,\td,\sd,s,u] + \delta[t] \sum_{x^+} p[x^+ \mid x,b] \: V[x^+].
\end{align*}    
\end{subequations}
Here, $\delta[t]$ is a \emph{state-dependent discount factor} satisfying $\delta[t < \tend] = 1$ and $\delta[t = \tend] \in [0, 1)$, i.e., discounting occurs only on transitions to future days, but not within the same day.
Note that $\prod_t \delta[t] < 1$, and therefore the Bellman recursion specifying $\bm{V}(\bm d,\bm \pi)$ is \emph{eventually contracting} and has a unique, continuous solution in $(\bm d,\bm \pi)$~\cite{stachurski2021dynamic}.

\begin{definition}
A social state $(\bm{d^*},\bm{\pi^*})$ is a \emph{\acrfull{SNE}} if, for all $x = [t,\tilde x] \in \X$,
\begin{subequations}
\begin{align}
    d^*[\tilde x \mid t] &= \sum_{\tilde x^-} d^*[\tilde x^- \mid t^-] \: P[t,\tilde x \mid t^-,\tilde x^-], \label{eq:SNE-1} \\
    \bm{\pi^*}[\cdot \mid x] &\in \argmax_{\sigma \in \Delta(\B[\ell,s,k])} \sum_b \sigma[b] \: Q[x,b], \label{eq:SNE-2} 
\end{align}
\end{subequations}
where $t^-=t - \Delta t$ for $t > \tstart$ and $t^- = \tend$ for $t=\tstart$.
\end{definition}

The existence of a \gls{SNE} is guaranteed for general \gls{DPG}s if the immediate payoff and state transition functions are continuous in the social state $(\bm d,\bm \pi)$~\cite{elokda2024dynamic}.
Due to the continuous construction of these functions using~\eqref{eq:auction-outcome-continuous}, and that continuity is not lost with the state-dependent discounting scheme, the following theorem is a consequence of~\cite[Proposition~1]{elokda2024dynamic}.

\begin{theorem}
\label{th:SNE-exists}
A \gls{SNE} is guaranteed to exist in the \gls{EV} charging karma economy with immediate payoff function~\eqref{eq:immediate-rewards} and state transition function~\eqref{eq:state-transitions}.
\end{theorem}

%% file: sections/numerical.tex
\section{Numerical Analysis}
\label{sec:numerical}

We now perform a numerical analysis of the \gls{SNE} of the \gls{EV} charging karma economy, computed using the evolutionary dynamics-inspired algorithm introduced in~\cite{elokda2024dynamic,elokda2024self}.
Due to the size of the state space in this complex setting, the \gls{SNE} computation was rather expensive, and we terminated the algorithm after 1000 iterations, which required approximately 48hrs to complete with a quad-core Intel i7-8550U CPU and 16GB RAM machine.

\subsection{Settings and Parameters}
\label{sec:params}
We consider two representative settings in our numerical analysis: \emph{moderate scarcity} and \emph{high scarcity}.
Both settings consider a single charging station, e.g., workplace charging, with nominal charging rate $\enom=8$kWh (corresponding to level-2 AC charging).
The average capacity is $8 / 3$kWh / user in moderate scarcity, and 2kWh / user in high scarcity, i.e., the charging station can admit at most $1 / 3$ or $1 / 4$ of the total population to charge at a time, respectively.
In both settings, we consider a 12-hour time period with $\tstart=7$:00, $\tend=19$:00, $\Delta t=1$; maximum \gls{SOC} $\smax=64$kWh (corresponding to $80\%$ of 80kWh battery); system average karma $\kbar=9$ and maximum karma $\kmax=18$; continuous approximation parameter $\epsilon = 10^{-4}$; and end-of-day discount factor $\delta[\tend]=0.99$.
The arrival time distribution follows typical workplace patterns with $\Prob^\textup a[t] = 0.25$ for $t \in \{\textup{7:00}, \textup{8:00}, \textup{9:00}\}$ and $\Prob^\textup a[t] = 0.125$ for $t \in \{\textup{10:00}, \textup{11:00}\}$.
The demand state distribution is $\phi[\td,\sd,u] = \Prob[\td] \, \Prob[\sd] \, \Prob[u]$, with each day's trip deadline, desired \gls{SOC} and urgency drawn independently from the following distributions: $\Prob[\td=8\textup{hr}]=0.75$ (default deadline), $\Prob[\td=4\textup{hr}]=0.25$ (occasional short deadline); $\Prob[\sd=32\textup{kWh}]=0.75$ (default $40\%$ charge requirement), $\Prob[\sd=48\textup{kWh}]=0.2$ (occasional $60\%$ charge requirement), $\Prob[\sd=64\textup{kWh}]=0.05$ (rare $80\%$ charge requirement); and $\Prob[u=1]=0.75$ (default urgency), $\Prob[u=9]=0.25$ (occasional high urgency).
Finally, we suppose that \gls{EV} users do not have access to external charging and thus the end-of-day \gls{SOC} transitions are deterministic with $s^\textup{start}=\max\left\{s^\textup{end} - \sd + \enom, \enom\right\}$\footnote{This corresponds to maintaining a minimum \gls{SOC} of $\enom$ ($10\%$ of battery capacity) and thus the actual trip consumption is $\sd - \enom$.}.

\subsection{Performance Measures and Benchmarks}

We consider two performance measures: the per-day \emph{average wait time beyond deadline} $\bar w(\bm{d^*}, \bm{\pi^*})$ (or average wait time for short) and \emph{average payoff} $\bar r(\bm{d^*}, \bm{\pi^*})$, given respectively at \gls{SNE} $(\bm{d^*}, \bm{\pi^*})$ as
\begin{align}
    \bar w(\bm{d^*}, \bm{\pi^*}) &= \sum_{t,\sd,u,s,k} d^*[\ell=1,\td=0,\sd,u,s,k \mid t], \label{eq:wait} \\
    \bar r(\bm{d^*}, \bm{\pi^*}) &= \sum_{t,\tilde x} d^*[\tilde x\mid t] \, r[t,\tilde x]. \label{eq:average-payoff}
\end{align}
The average wait time as per~\eqref{eq:wait} is defined as the long-run fraction of times spent in ``waiting'' states $[\ell=1,\td=0]$ at which the deadline has passed but the \gls{EV} must still remain in the charging station.
The average payoff as per~\eqref{eq:average-payoff} is the long-run average of the immediate payoff function~\eqref{eq:immediate-rewards} with respect to the stationary distribution $\bm{d^*}$, and additionally takes into account the urgency in waiting states.

Moreover, we consider two popular benchmarks in the scheduling literature: \gls{FCFS} and \gls{EDF}.
Under \gls{FCFS}, \gls{EV} users are prioritized based on order of arrival and remain charging until they either reach $\smax$ or depart from the charging station.
This baseline scheme is agnostic to the private demand states $[\td,\sd,u]$.
Under \gls{EDF}, \gls{EV} users report their deadlines $\td$ and desired \gls{SOC} $\sd$, and those with earliest $\td$ gain priority (unless they have already reached $\sd$).
\gls{EDF} is considered the workhorse scheduler in deadline-sensitive contexts since it guarantees meeting all deadlines whenever feasible~\cite{chetto1989some,ahmad2014earliest}.
However, in addition to assuming truthful reporting of $\td$ and $\sd$, it does not take the urgency $u$ into account.
Each of these two benchmarks induces a Markov chain on the states whose derivation is similar to Section~\ref{sec:state-transitions}\footnote{The main difference to Section~\ref{sec:state-transitions} is that priority to charge, i.e., Equation~\eqref{eq:auction-outcome-continuous}, is based on arrival times or deadlines instead of bids.}.
The performance measures of each benchmark are computed analogously to Equations~\eqref{eq:wait}--\eqref{eq:average-payoff} at the stationary distribution of the respective Markov chain.

\subsection{Results}

\begin{figure}[t]
    \centering
    \begin{subfigure}[b]{0.23\textwidth}
        \includegraphics[width=\textwidth]{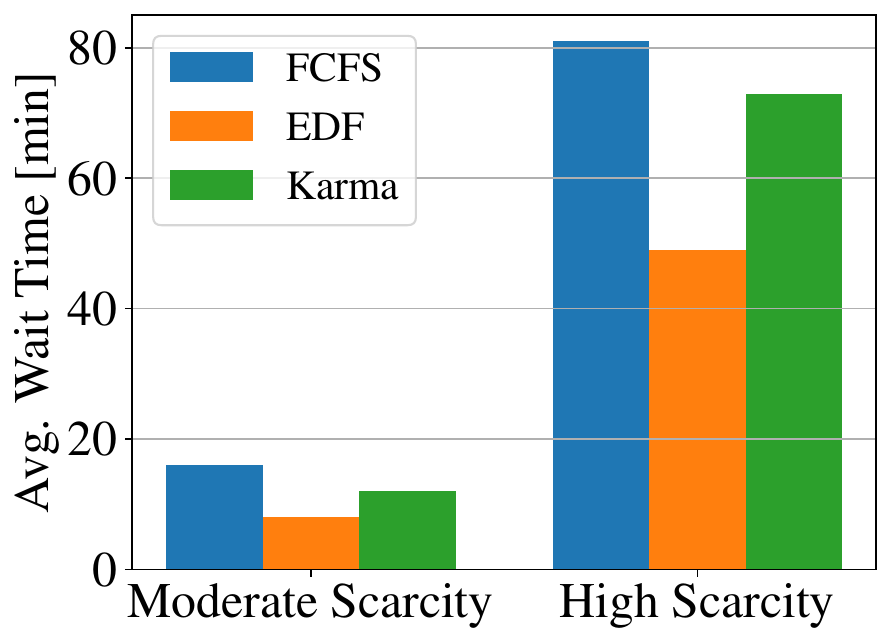}
        \caption{Average wait time $\bar w$}
        \label{fig:average-waiting-time}
    \end{subfigure}
    \hfil
    \begin{subfigure}[b]{0.23\textwidth}
        \includegraphics[width=\textwidth]{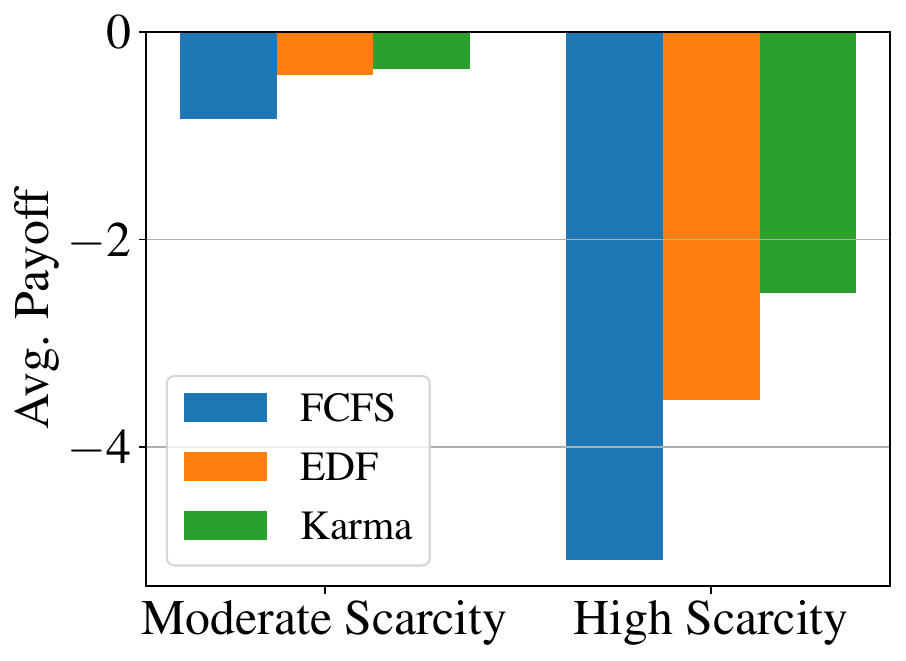}
        \caption{Average payoff $\bar r$}
        \label{fig:average-payoff}
    \end{subfigure}

    \caption{Performance of karma versus \gls{FCFS} and \gls{EDF}.}
    \label{fig:performance}
\end{figure}

\begin{figure}[t]
    \centering
    \begin{subfigure}[b]{0.23\textwidth}
        \includegraphics[width=\textwidth]{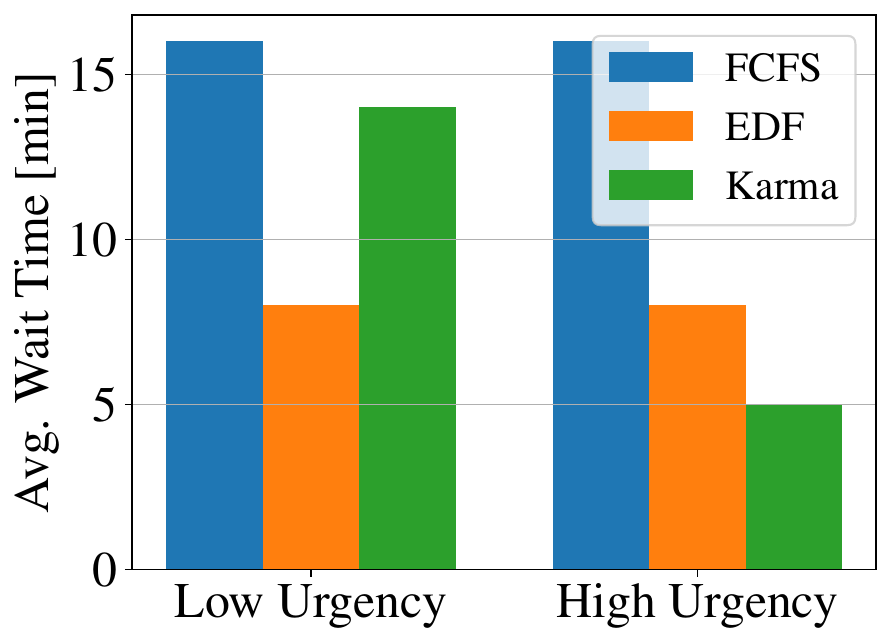}
        \caption{Moderate Scarcity}
        \label{fig:moderate-scarcity-wait-per-urgency}    
    \end{subfigure}
    \hfil
    \begin{subfigure}[b]{0.23\textwidth}
        \includegraphics[width=\textwidth]{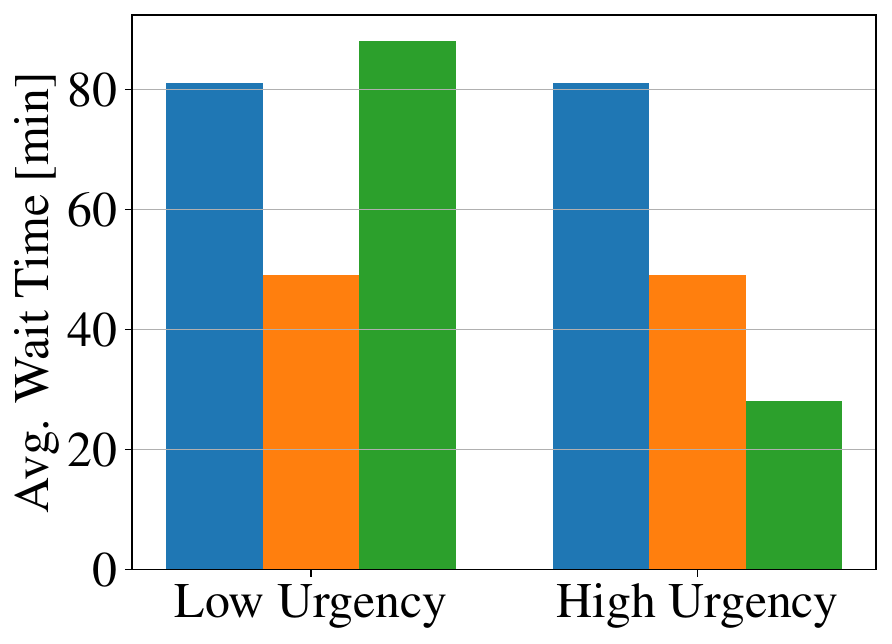}
        \caption{High Scarcity}
        \label{fig:high-scarcity-wait-per-urgency}
    \end{subfigure}

    \caption{Average wait time conditional on urgency $\bar w[u]$.}
    \label{fig:wait-per-urgency}
\end{figure}

We first show the performance measures of the karma scheme versus the benchmarks for both settings of moderate and high scarcity in Fig.~\ref{fig:performance}. 
The average wait time is highest for \gls{FCFS}, lowest for \gls{EDF}, and the karma scheme lies in between these two benchmarks, cf. Fig.~\ref{fig:average-waiting-time}.
Higher average wait time indicates that the karma scheme is susceptible to miss more deadlines than \gls{EDF} under the assumption of truthful reporting in \gls{EDF}.
Nonetheless, the karma scheme attains the maximum average payoff, cf. Fig.~\ref{fig:average-payoff}, owing to its ability to tailor to the private urgency in addition to the private deadline. As illustrated in Fig.~\ref{fig:wait-per-urgency}, once we condition the average wait time on the urgency state\footnote{For urgency $u$, $\bar w[u](\bm d^*, \bm \pi^*)=\frac{\sum_{t,\sd,s,k} d^*[\ell=1,\td=0,\sd,u,s,k \mid t]}{\Prob[u]}$.}, karma drastically reduces the wait times of high urgency users compared to \gls{EDF}. 
Karma thus effectively balances between meeting deadlines and prioritizing high urgency trips.

\begin{figure}[t]
    \centering
    \begin{subfigure}[b]{0.23\textwidth}
        \includegraphics[height=0.77\textwidth]{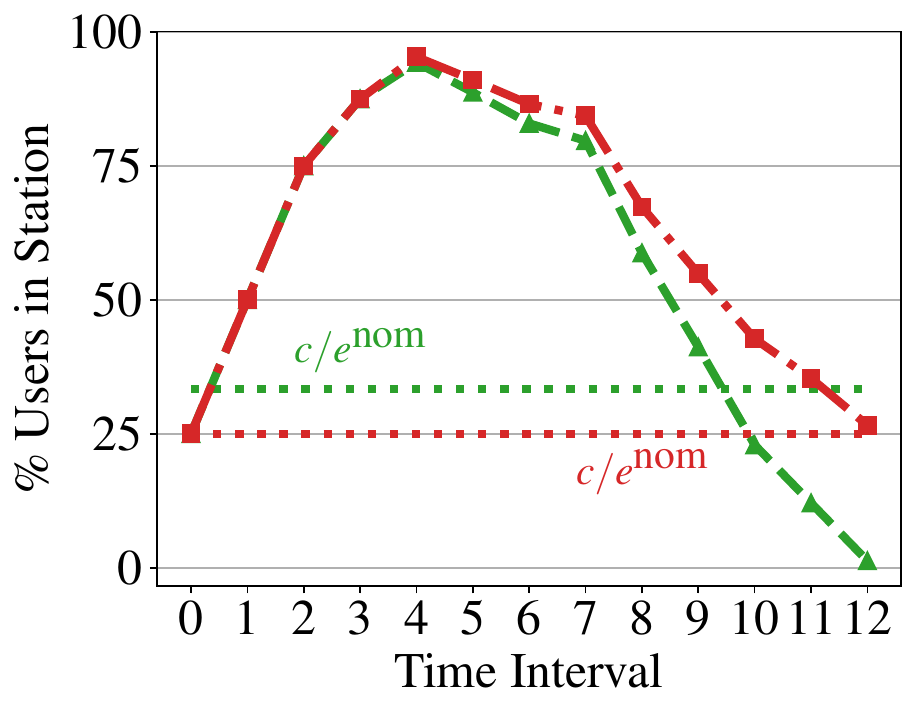}
        \caption{$\Prob[\ell=1 \mid t]$}
        \label{fig:presence-per-time}    
    \end{subfigure}
    \hfil
    \begin{subfigure}[b]{0.23\textwidth}
        \includegraphics[height=0.75\textwidth]{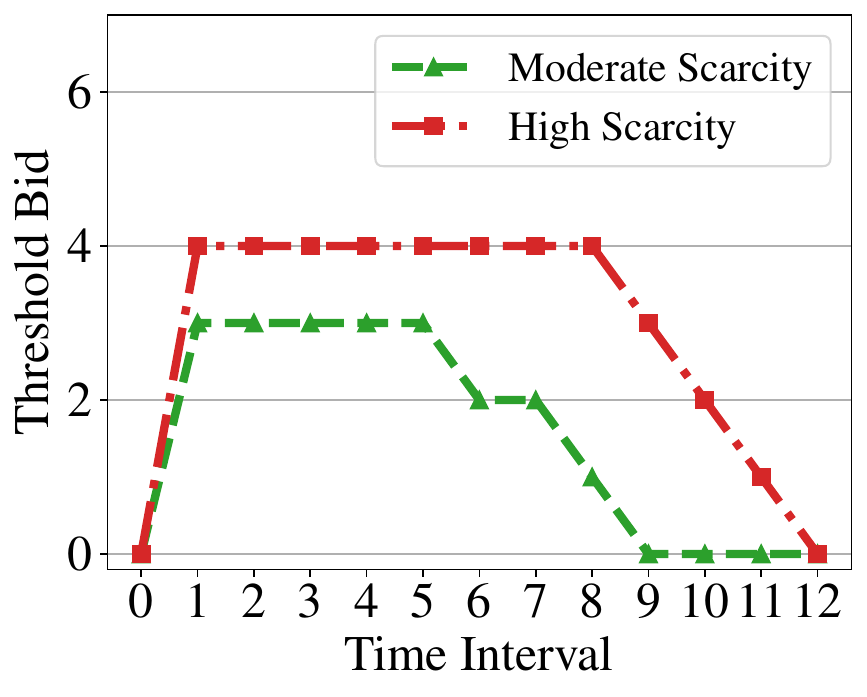}
        \caption{$b^*[t]$}
        \label{fig:bstar-per-time}
    \end{subfigure}

    \caption{Charging station occupancy and threshold bids.}
    \label{fig:bidding}
\end{figure}

Finally, we focus on the competitive landscape under the karma scheme.
Fig.~\ref{fig:presence-per-time} shows the fraction of users present in the charging station over time, defined as $\Prob[\ell=1 \mid t](\bm{d^*},\bm{\pi^*})=\sum_{\td,\sd,u,s,k}d^*[\ell=1,\td,\sd,u,s,k\mid t]$, as well as the fraction that can be admitted to charge in each time interval $c / \enom$.
Fig.~\ref{fig:bstar-per-time} shows the corresponding threshold bid $b^*[t]$ to be among the $c / \enom$ highest bidders.
As expected, in the high scarcity setting, there is prolonged congestion of the charging station, which in turn results in higher threshold bids than in the moderate scarcity setting.
The threshold bids decay to zero towards the end of the day as the congestion gets alleviated.

%% file: sections/conclusion.tex
\section{conclusion}
\label{sec:conclusion}

This paper introduces a karma-based scheme for flexible \gls{EV} charging, framing the system as a \gls{DPG} with a guaranteed \gls{SNE}.
Numerical analysis demonstrates that the karma scheme outperforms benchmark scheduling schemes (\gls{FCFS} and \gls{EDF}) in terms of average payoffs experienced by the users.
This superiority stems from the mechanism's unique capacity to align allocation with both heterogeneous deadlines and private urgency.
While this study assumes that users can participate in relatively frequent bidding, we envision automating the bidding based on the \gls{SNE} policy in the future.
It is also important to consider user heterogeneity, e.g., heterogeneous battery capacities, charging rates, access to home charging, commuting patterns, etc..
With heterogeneous users, previous work has shown that karma economies maximize \gls{LNW} under some assumptions that do not hold in the present setting (e.g., that utility for resources is additively separable)~\cite{elokda2025vision}, and it will be interesting to extend this result.
Another exciting research direction is to investigate the added benefit of coupling \gls{EV} charging with other resource allocations, e.g., traffic and household energy.